\documentclass[superscriptaddress, reprint]{revtex4-1}
\usepackage{multirow}
\usepackage{hyperref}
\usepackage[version=4]{mhchem} 
\usepackage{graphicx}
\usepackage{chemmacros}
\usepackage{gensymb}
\usepackage{dcolumn}
\usepackage{longtable}
\usepackage{array}
\usepackage{amssymb}

\begin{document}

\title{Spontaneous rotation of ferrimagnetism driven by antiferromagnetic spin canting}

\author{A. M. Vibhakar}
\affiliation{Clarendon Laboratory, Department of Physics, University of Oxford, Oxford, OX1 3PU, United Kingdom}
\author{D. D. Khalyavin}
\affiliation{ ISIS facility, Rutherford Appleton Laboratory-STFC, Chilton, Didcot, OX11 0QX, United Kingdom}
\author{P. Manuel}
\affiliation{ ISIS facility, Rutherford Appleton Laboratory-STFC, Chilton, Didcot, OX11 0QX, United Kingdom}
\author{J. Liu}
\affiliation{Clarendon Laboratory, Department of Physics, University of Oxford, Oxford, OX1 3PU, United Kingdom}
\author{A. A. Belik}
\affiliation{Research Center for Functional Materials, National Institute for
Materials Science (NIMS), Namiki 1-1, Tsukuba, Ibaraki 305-0044, Japan}
\author{R. D. Johnson}
\affiliation{Clarendon Laboratory, Department of Physics, University of Oxford, Oxford, OX1 3PU, United Kingdom}
\affiliation{Department of Physics and Astronomy, University College London, Gower Street, London, WC1E 6BT, United Kingdom}
\date{\today}

\begin{abstract}
Spin-reorientation phase transitions that involve the rotation of a crystal's magnetisation have been well characterised in distorted perovskite oxides such as the orthoferrites. In these systems spin reorientation occurs due to competing rare earth and transition metal anisotropies coupled via $f-d$ exchange. Here, we demonstrate an alternative paradigm for spin reorientation in distorted perovskites. We show that the $R_2$\ce{CuMnMn4O12} ($R$ = Y or Dy) triple A-site columnar ordered quadruple perovskites have three ordered magnetic phases and up to two spin reorientation phase transitions. Unlike the spin reorientation phenomena in other distorted perovskites, these transitions are independent of rare earth magnetism, but are instead driven by an instability towards antiferromagnetic spin canting likely originating in frustrated Heisenberg exchange interactions, and the competition between Dzyaloshinskii-Moriya and single-ion anisotropies. 
\end{abstract}
 
\maketitle

Temperature induced spin-reorientation (SR) phase transitions occur between two long-range ordered magnetic phases \cite{1968Horner, 1976Belov, 1997Chikazumi}, and can involve the rotation of the staggered magnetisation in antiferromagnets \cite{2010Marcinkova, 2010Kimber, 2015Corkett, 2015Zhang}, the net magnetisation in ferromagnets and ferrimagnets \cite{1991Tie-song, 1968Houghton}, or both coupled together \cite{2018Bolletta}. The discovery of SR transitions and the establishment of their microscopic origin has attracted substantial attention within the condensed matter physics community both in the fundamental study of order and dynamics in quantum magnetism, and in more applied directions where magnetisation switching is an essential ingredient for realising nanoscale components that form the building blocks of spintronic based technology \cite{2018Fita}, such as those used in spin-torque devices \cite{2013Locatelli}. 

SR-based devices lend themselves to operating in the ultrafast regime when there exists two antialigned magnetic sublattices, in for example antiferromagnets or ferrimagnets \cite{2009Kimel}, which is critical to achieving operational speeds that are comparable to those of charge-based devices \cite{2017Yang}. The rare-earth (RE) orthoferrite \cite{1956Geller, 1960Koehler} distorted-perovskite oxides (generic chemical formula \ce{ABO3}, A = RE and B = Fe) are archetypal examples of materials that host ultrafast SR phase transitions \cite{2004Kimel}. Antiferromagnetic (AFM) order resides on the transition metal (TM) sublattice at all temperatures below $T_N$, and in certain symmetries a weak ferromagnetic (FM) moment appears through spin canting, which is coupled to the primary AFM order via the Dzyaloshinskii-Moriya interaction \cite{1962Treves}. For compounds with a magnetic RE species, $f-d$ exchange interactions have been found to couple either the weak FM moment or the staggered magnetisation of the TM sublattice to the RE sublattice. A temperature induced SR transition can then result if the RE and TM ions have competing magnetic anisotropies \cite{1969White}. 

A SR phase transition driven by the same mechanism of $f-d$ coupled competing magnetic anisotropies was recently observed in the triple A-site columnar-ordered quadruple perovskite manganite \ce{TmMn3O6} \cite{2019Vibhakar} (generic chemical formula \ce{$A$2$A'A''$B4O12}, $A$ = Tm and $A'$, $A''$, $B$ = Mn). This material belongs to a relatively understudied family of distorted perovskites related to the orthoferrites, but with a different octahedral tilt pattern of $a^+a^+c^-$ instead of $a^-a^-c^+$ (in Glazer notation \cite{1972Glazernotation}), which affords a complex ordering of $A$, $A'$ and $A''$ cations \cite{2018BelikTriple}. \ce{TmMn3O6} adopts a ferrimagnetic (FIM) structure and boasts an uncompensated magnetisation of approximately $1\mu_\mathrm{B}$ per TM \cite{2019Vibhakar}, as opposed to the weak FM canting of typically $\leq$0.1 $\mu_\mathrm{B}$ per TM found in the orthoferrites \cite{1958Bozorth}. Hence, a much larger magnetisation rotates at the SR transition in \ce{TmMn3O6}, giving, for example, the potential for a stronger exchange bias in device heterostructures whilst maintaining ultrafast switching potential.

In this paper, we demonstrate a new paradigm for SR transitions in distorted perovskites. We show that both \ce{Y2CuMnMn4O12} (YCMO) and \ce{Dy2CuMnMn4O12} (DCMO) columnar-ordered quadruple perovskites adopt a canted FIM structure ($m||b$) below $T_1$ = 175 K and 159 K, and undergo a SR transition to a collinear FIM ($m||c$) at $T_2$ = 115 K and 125 K, respectively. Both materials undergo another phase transition at $T_3$ = 17 K, to a canted FIM ground state with a doubled magnetic unit cell. In YCMO, this transition is accompanied by a rotation of the magnetisation back to $m||b$. In DCMO SR does not occur at $T_3$, owing to an additional Ising-like anisotropy of the polarised Dy$^{3+}$ crystal electric field (CEF) ground state. We show that, unlike other distorted perovskites, the observed SR phenomena in YCMO and DCMO is independent of RE magnetism, but can instead be driven by an instability towards AFM spin canting combined with Dzyaloshinskii-Moriya and single-ion anisotropies. 

\begin{figure}
\includegraphics[width=\linewidth]{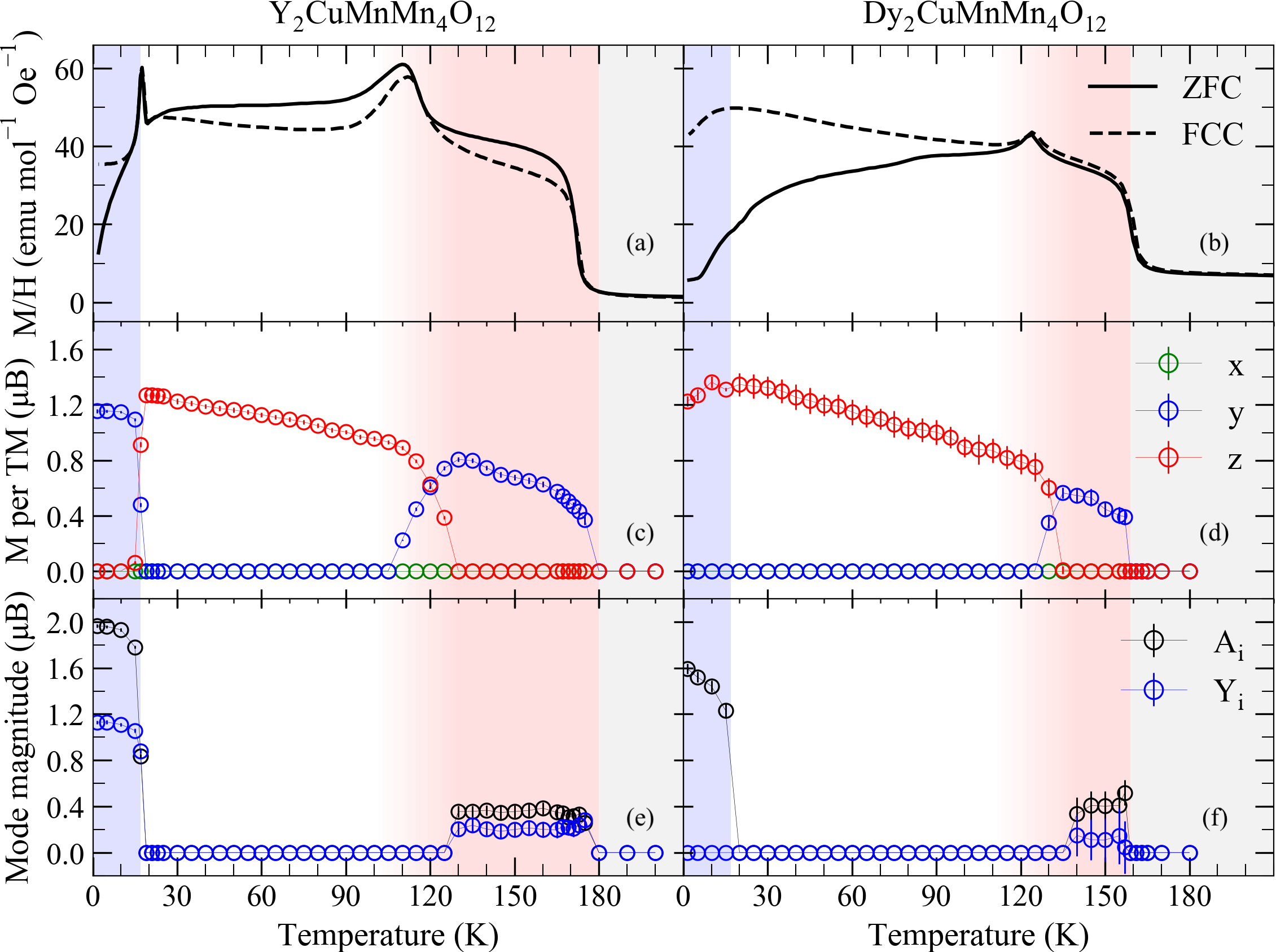}
\caption{Temperature dependence of (a)-(b) ZFC and FCC magnetisation measurements under an applied DC field of 100 Oe (c)-(d) the net magnetisation per TM. The net magnetisation was calculated by taking the sum over all magnetic sites in the unit cell (Cu1 had a contribution of zero to the net magnetisation) and (e)-(f) the Mn3 moment magnitudes for each AFM spin canting mode.}
\label{FIG::tempdep}
\end{figure}

\begin{figure}
\includegraphics[width=\linewidth]{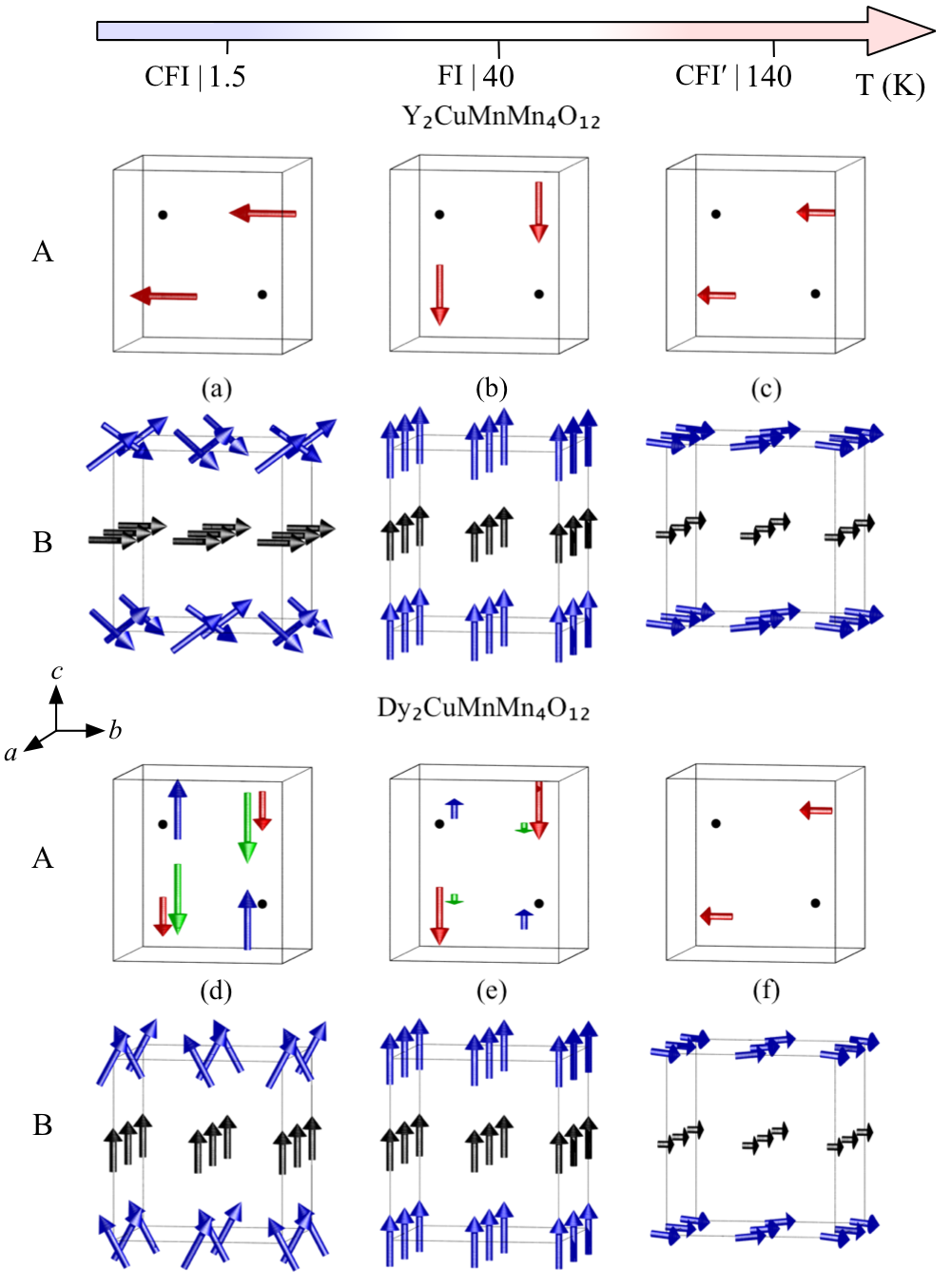}
\caption{The refined magnetic structures, separated into the A and B sites, of (a)-(c) YCMO and (d)-(f) DCMO at temperatures of 1.5 K, 40 K and 140 K. For the A sites, R1 is shown in blue, R2 in green, Cu1 in black and Mn2 in red. For the B sites Mn3 is shown in blue and Mn4 in black. The crystallographic unit cell is drawn in black. Note that at 1.5 K the propagation vector of the spin canting is (0,0,$\frac{1}{2}$) and thus (a) and (d) should be doubled along $c$ to represent the full magnetic unit cell.}
\label{FIG::magneticstructure}
\end{figure}

A 2.06g polycrystalline sample of YCMO was prepared from a stoichiometric mixture of h-\ce{YMnO3}, \ce{Mn2O3}, \ce{CuO} and \ce{MnO_{1.839}}. The mixture was annealed in small batches at 6 GPa and $\sim$1670 K for 2 h in Pt capsules in a belt-type high-pressure apparatus, before the mixture was quenched to room temperature. A 1.32g polycrystalline sample of DCMO was prepared in the same way, using \ce{Dy2O3} instead of the h-\ce{YMnO3} precursor \cite{2018BelikTriple}. DC magnetization measurements were performed using a SQUID magnetometer (Quantum Design, MPMS-XL-7T) between 2 and 400 K under both zero-field-cooled (ZFC) and field-cooled on cooling (FCC) conditions. Time of flight neutron diffraction measurements were performed using WISH at ISIS \cite{2011ChaponWISH}. Both samples were cooled to 1.5 K and neutron diffraction data were collected on warming through all magnetic phases in 5 K steps (2 K steps were used close to $T_1$ and also close to $T_3$ for YCMO). In addition, data were collected with high counting statistics at temperatures representative of each magnetic phase. The crystal structure for both samples was refined against data collected in the paramagnetic phase using the published $Pmmn$ structure \cite{2018BelikTriple} (full details are given in the Supplemental Material (SM) Sec. S1 \footnote{See Supplemental Material for further details of the neutron powder diffraction refinement, the symmetry analysis, the Dy$^{3+}$ crystal electric field calculations and the results of the ESR measurements, which includes Refs. \cite{2017Zhang, 1991Brese, 2006Isodistort, 1993Rodriguez, 1964Hutchings, 1979Freeman, 2000Ivanshin, 1999Koksal}}). The two $A$ sites (labeled R1 and R2) were found to be fully occupied by Y$^{3+}$ and Dy$^{3+}$ in each of the samples. The $A'$ sites (labeled Cu1) were occupied by 79\% Cu$^{2+}$ and 21\% Mn$^{3+}$ in YCMO, and 70\% Cu$^{2+}$ and 30\% Mn$^{3+}$ in DCMO. The $A''$ sites (labeled Mn2) were occupied by 74\% Mn$^{2+}$ and 26\% Y$^{3+}$ in YCMO, and 78\% Mn$^{2+}$ and 22\% Dy$^{3+}$ in DCMO. The two symmetry inequivalent B sites form layers stacked alternately along $c$, labeled Mn3 and Mn4. The Mn3 layers were fully occupied by Mn$^{3+}$ in YCMO, but had a 9\% occupation of Cu$^{2+}$ in DCMO. The Mn4 layers were fully occupied by Mn$^{4+}$ in both samples. In the following we present neutron powder diffraction data analysed in terms of symmetry-adapted magnetic modes, $F_i$, $A_i$, $X_i$, and $Y_i$ (defined in the SM \cite{Note1} for the A and B sublattices), where $i$ represents the direction of the moment components in the basis $x||a$, $y||b$, $z||c$.

\begin{figure*}
\includegraphics[width=\linewidth]{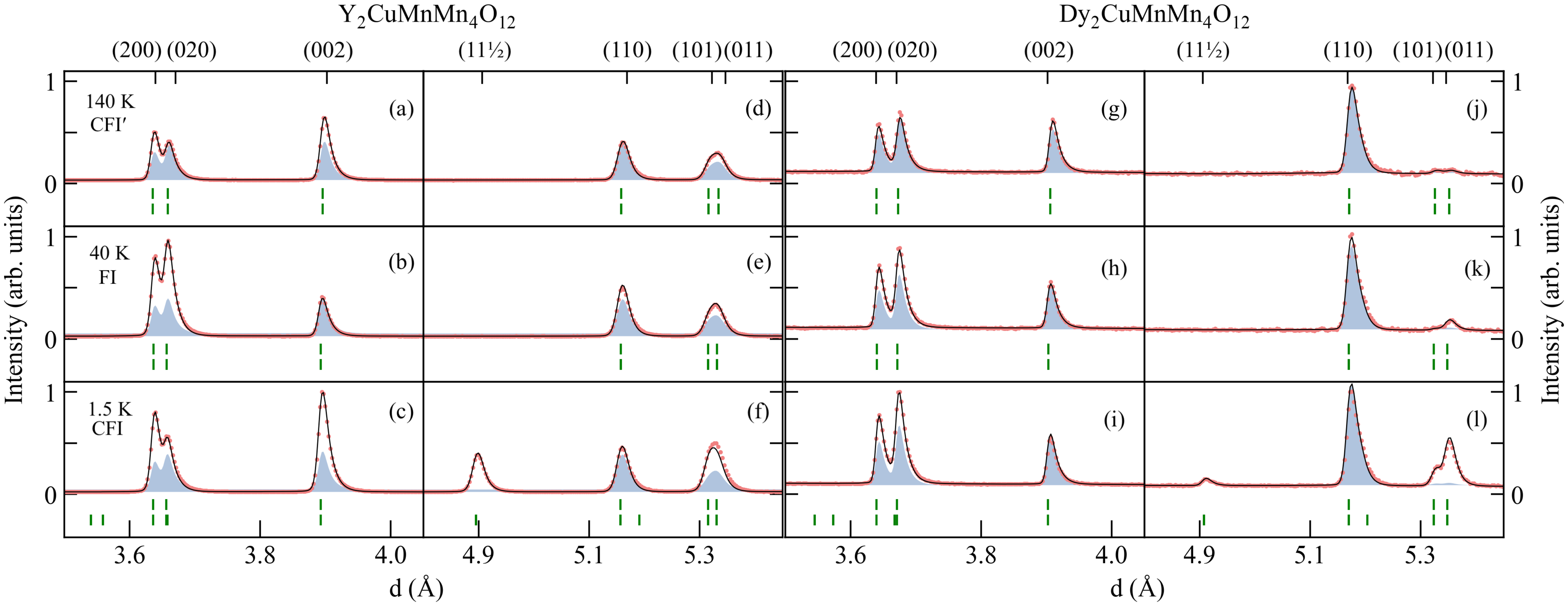}
\caption{Neutron powder diffraction data taken on WISH; (a), (d), (g) and (j) were collected at 140 K; (b), (e), (h) and (k) at 40 K and (c), (f), (i) and (l) at 1.5 K. (a)-(c) \{200\} and (d)-(f) \{110\}  reflections from YCMO, and (g)-(i) \{200\} and (j)-(l) \{110\} reflections from DCMO. The data are shown by the red circles, the fit by the solid black line and the blue shading represents the nuclear contribution to scattered intensity, including an impurity phase for DCMO. The green tick marks indicate, from top to bottom, the nuclear and magnetic reflections.}
\label{FIG::neutrondiffraction}
\end{figure*}

At $T_1$ = 175 K new $\Gamma$-point intensities appeared in the diffraction pattern for YCMO (Fig. S4 \cite{Note1}) that occurred concomitantly with a rapid increase in the DC magnetic susceptibility shown in Figure \ref{FIG::tempdep}a, which together indicate the development of long range FM sublattices. Rietveld refinement of the diffraction data showed that in this first magnetic phase (labeled CFI$'$) Mn2 and Mn4 sites order with $F_y$ modes, and the Mn3 sites order with $A_x, F_y, Y_z$ modes (Figure \ref{FIG::magneticstructure}c). Key features of the magnetic structure can be directly observed in the diffraction pattern. For example the \{200\} family of magnetic reflections, Figures \ref{FIG::neutrondiffraction}a-c, originate in scattering from A- and B-site $F_i$ modes only, and their relative intensities provide a direct measure of the direction, $i$, of the respective moments. Zero magnetic intensity at (020) and equal magnetic intensities at (200) and (002), as shown in Figure \ref{FIG::neutrondiffraction}a, is uniquely consistent with $F_y$ modes. The \{110\} family of magnetic intensities shown in Figure \ref{FIG::neutrondiffraction}d-f are also sensitive to $F_i$ modes at the A-sites, and AFM $A_i$, $X_i$, $Y_i$ modes (instead of $F_i$ modes) at the B-sites. In this phase zero magnetic intensity at (110), and equal magnetic intensities at (011) and (101), Figure \ref{FIG::neutrondiffraction}d, can only be accounted for by the presence of $A_x, Y_z$ modes on Mn3 layers in addition to the Mn2 $F_y$ mode. The $F_y$-modes combine to give a FIM structure with a net magnetisation parallel to $y$, and the $A_x$ and $Y_z$ modes give rise to spin canting of the Mn3 layers with the same amplitude in every Mn3 layer. The temperature dependence of the net magnetisation arising from the $F_y$ modes, and the perpendicular moment magnitude of the $A_x$ and $Y_z$ canting are given in Figure \ref{FIG::tempdep}c and e, respectively. As the temperature is lowered the magnetisation increases, in agreement with the temperature dependence of all symmetry inequivalent magnetic moments (Fig. S4 \cite{Note1}).

At $T_2$ = 115 K the shift of magnetic intensity from the (002) to the (020) reflection, whilst the (200) intensity remains the same (Figure \ref{FIG::neutrondiffraction}b), indicates that the $F_i$ mode, and therefore the net magnetisation, rotated from the $b$ axis to the $c$ axis. This SR transition is also apparent in magnetometry measurements shown in Figure \ref{FIG::tempdep}a, that feature a characteristic cusplike increase in the magnetisation \cite{2019Vibhakar}. Rietveld refinement of data measured at 40 K showed that this second magnetic phase (labeled FI) has a collinear FIM structure where the Mn2, Mn3 and Mn4 sublattices order with $F_z$ modes (Figure \ref{FIG::magneticstructure}b). The SR transition at $T_2$ was found to occur over a 20 K interval, as shown in Figure \ref{FIG::tempdep}a and c. The collinear phase then persists on cooling to a second SR transition at $T_3$ = 17 K. 

Below $T_3$, two significant changes to the diffraction pattern were observed. Firstly, eight new magnetic peaks appeared that could be indexed with a \textbf{k} = (0,0,0.5) propagation vector (Z-point). Secondly, intensity shifted from the (020) to the (002) reflection, showing that the $F_i$ modes rotated back to the $b$ axis. In this third magnetic phase (labeled CFI), Mn2, Mn3 and Mn4 $F_y$ modes accounted for all $\Gamma$-point magnetic intensities apart from a small intensity at (011) that is likely due to the presence of an unidentified magnetic impurity. The magnetic peaks at the Z-point were fitted with $Y_x, A_z$ AFM canting modes on the Mn3 layers, the amplitude of which changes sign from one Mn3 layer to the next (Figure \ref{FIG::magneticstructure}a). We note that in all magnetic phases our analysis indicated zero moment on the Cu1 sites, despite strong exchange coupling of Cu demonstrated by ESR measurements (Fig. S5 \cite{Note1}). This effect could occur if the minority Mn$^{3+}$ and majority Cu$^{2+}$ ions, randomly located on the Cu1 site, had antiparallel moments, giving an average moment lower than the sensitivity of our experiment \footnote{This scenario is akin to alloys of niobium-vanadium and compounds with a mixture of Deuterium and Hydrogen, that give zero nuclear scattering length}.

In DCMO, the first two magnetic phases, CFI$'$ and FI (Figures \ref{FIG::magneticstructure}f and \ref{FIG::magneticstructure}e), were found to be identical to YCMO, except that in the FI phase an additional polarised moment developed on the Dy1 and Dy2 sublattices with $m||c$ \cite{Note1}. CEF calculations (Sec. S5 \cite{Note1}) showed that the lowest lying Kramers doublet of both Dy1 and Dy2 is well separated from excited states, and has a strong Ising like g-tensor anisotropy $||z$, which is consistent with the onset of a Dy1 and Dy2 moment only below the SR transition at $T_2$ = 125 K. Evidence for the polarised Dy1 and Dy2 moments was found at the (011) and (101) reflections (Figure \ref{FIG::neutrondiffraction}k), which are expected to be of equal intensity in the absence of a moment on the RE sites, as was the case for the YCMO magnetic structure refined at 40 K (Figure \ref{FIG::neutrondiffraction}e), but instead had unequal intensities consistent with a finite moment on the Dy sublattices. By comparison with YCMO, it is apparent that the additional Dy magnetism had two significant effects on the behaviour of the system. Firstly the SR transition between CFI$'$ and FI became sharper, Figure \ref{FIG::tempdep}b and \ref{FIG::tempdep}d, and secondly at $T_3$ = 17 K it prevented the SR between phases FI and CFI (as indicated by the absence of any changes at the \{200\} family of reflections (Figure \ref{FIG::neutrondiffraction}i), and the absence of a cusp in the magnetic susceptibility, Figure \ref{FIG::tempdep}b). Instead the CFI phase of DCMO had $\Gamma$-point $F_z$ modes across all sites that maintained the FIM structure with magnetisation $||c$ (Figure \ref{FIG::magneticstructure}d). A Z-point peak also appeared in the CFI phase, which could be fit by a single AFM canting mode on the Mn3 layer (either $A_x$ or $A_y$).

Spin canting typically arises when the Dzyaloshinskii-Moriya (DM) interaction is allowed by symmetry (\emph{e.g.} in orthoferrite YFeO$_3$ DM exchange induces a $\sim$0.3$^\circ$ canting of Fe spins \cite{1967Eibschutz}, which is small due to the DM interaction being approximately thirty times weaker than Heisenberg exchange \cite{2018Amelin}). By comparison, Mn3 spins in $R_2$\ce{CuMnMn4O12} exhibit canting angles of up to $\sim$30$^\circ$ (Fig. S4 \cite{Note1}), which are improbably large for DM-induced spin canting. Furthermore, the symmetry of spin canting found in the CFI phase of DCMO is incompatible with DM exchange (as explained in the next paragraph). Instead, large spin canting in $R_2$\ce{CuMnMn4O12} can be established by the competition between effective FM (Mn3-O-A-O-Mn3) and AFM (Mn3-O-Mn3) isotropic Heisenberg exchange between NN Mn3 spins in a given layer, as described in reference \citenum{2019Vibhakar}. The former interactions favour $F_i$ modes, whereas the latter primarily favour an $A_i$ mode, which can be augmented by $X_i$ or $Y_i$ modes in the case of inequivalent exchange along $x$ and $y$. In the mean field approximation one can show that this competition leads to two separate orders for the Mn3 sublattice; collinear FM or canted FM, which can be selected by tuning the ratio of the effective FM and AFM interaction strengths. This tuning is apparent when comparing the magnetic structures of $R_2$\ce{CuMnMn4O12} and \ce{TmMn3O6} \cite{2019Vibhakar}. In \ce{TmMn3O6} every $A'$ site was occupied by Mn$^{3+}$ and spin canting was not observed \cite{2019Vibhakar}. In $R_2$\ce{CuMnMn4O12} the magnetic moment on the $A'$ sites was considerably reduced by the substitution of Mn$^{3+}$ for Cu$^{2+}$, giving a smaller effective FM exchange and prevalent spin canting.

Whilst competing Heisenberg exchange can establish spin canting, it cannot alone determine the orientation of the magnetic moments with respect to the crystal structure. However, once the canted structure exists, DM interactions can play a dominant role in determining the spin anisotropy, as they will be optimized when the moment components of the primary magnetic modes ($F_i$, $A_i$, $Y_i$) are placed along specific crystallographic directions \cite{2015Khalyavin_1,2015Khalyavin_2}. Formally, DM interactions couple orthogonal moment components from two modes of different type (\emph{e.g.} $A_x, F_y$), both of which must transform by the same irreducible representation of the paramagnetic space group such that a bilinear free energy invariant exists \cite{1957Dzyaloshinsky}. In the SM (Table. S5) \cite{Note1} we tabulate linear combinations of modes categorised by their symmetry. It is apparent that the $A_x$, $F_y$, and $Y_z$ combination ($\Gamma_4^+$ irrep), as found in the CFI$'$ phase of both YCMO and DCMO, presents a unique combination of all three primary modes mutually coupled by DM interactions. In the CFI phase of YCMO, symmetry prohibits DM interactions between the $\Gamma$-point $F_y$ and the Z-point $A_i, Y_i$ modes. However the latter modes can couple via DM exchange in two symmetries: the $Z_1^+$ irrep ($Y_x, A_z$) and the $Z_4^+$ irrep ($A_x, Y_z$) (Table. S5 \cite{Note1}). Symmetry alone cannot determine which scenario is realised (the former was observed), but both involve moment components along $x$ and $z$ meaning that the net magnetisation from the $\Gamma$-point $F_i$ mode will be placed parallel to $b$ ($y$) to establish a constant moment ground state. To the contrary, in the CFI phase of DCMO the net magnetisation lies parallel to $c$ ($F_z$ modes). For $A_i$ or $Y_i$ modes to be included in a constant moment ground state they must have moment components perpendicular to $z$. However, there is no single irrep that transforms (couples) $A_x, Y_y$ or $Y_x, A_y$. Accordingly, the canted ground state of DCMO does not optimise antisymmetric exchange but is instead predominantly selected by the dominant, Ising-like anisotropy of Dy$^{3+}$. The intermediate FI phase of both YCMO and DCMO has an approximately collinear magnetic structure (within the sensitivity of the experiment), which drastically reduces the DM interactions. The $F_i$ modes then follow the weak easy axis single-ion-anisotropy of Mn$^{3+}$, which is parallel to the $c$ axis on average \cite{2015Whangbo}. 

In summary, \ce{Y2CuMnMn4O12} and \ce{Dy2CuMnMn4O12} host magnetic phase transitions between canted and collinear structures that are accompanied by spin reorientation from $m||b$ (favoured by DM in the canted phases) to $m||c$ (favoured by single ion anisotropy in the collinear phase), with the exception of the \ce{Dy2CuMnMn4O12} canted ground state where Dy$^{3+}$ Ising like anisotropy imposes $m||c$. Our results are in complete contrast to well established SR phenomena in other distorted perovskites, but instead demonstrate a mechanism for spin reorientation originating in spin canting instabilities. Importantly, these triple A-site columnar ordered quadruple perovskites have the potential to host ultrafast spin reorientation transitions whilst boasting a large 1$\mu B$ moment per TM, making them candidate materials to realise novel spin based technology. In future studies it would be interesting to investigate modification of the SR phenomena through further chemical substitution and doping on this triple A site columnar ordered perovskite framework. 

\begin{acknowledgments}
R.D.J. acknowledges support from a Royal Society University Research Fellowship. This study was supported in part by JSPS KAKENHI Grant No. JP16H04501, a research grant from Nippon Sheet Glass Foundation for Materials Science and Engineering (40-37), and Innovative Science and Technology Initiative for Security, ATLA, Japan.
\end{acknowledgments}

\end{document}